\def \AAP #1 #2 {{\em Astron. Astrophys.\/} {\bf #1}, #2}
\def \AAL #1 #2 {{\em Astron. Astrophys. Lett.\/} {\bf #1}, L#2}
\def \AAR #1 #2 {{\em Astron. Astrophys. Rev.\/} {\bf #1}, #2}
\def \AAS #1 #2 {{\em Astron. Astrophys. Suppl. Ser.\/} {\bf #1}, #2}
\def \AJ #1 #2 {{\em Astron. J.\/} {\bf #1}, #2}
\def \ANNREV #1 #2 {{\em Ann. Rev. Astron. Astrophys.\/} {\bf #1}, #2}
\def \APJ #1 #2 {{\em Astrophys. J.\/} {\bf #1}, #2}
\def \APJL #1 #2 {{\em Astrophys. J. Lett.\/} {\bf #1}, L#2}
\def \APJS #1 #2 {{\em Astrophys. J. Suppl.\/} {\bf #1}, #2}
\def \APSS #1 #2 {{\em Astrophys. Space Sci.\/} {\bf #1}, #2}
\def \ASR #1 #2 {{\em Adv. Space Res.\/} {\bf #1}, #2}
\def \BAIC #1 #2 {{\em Bull. Astron. Inst. Czechosl.\/} {\bf #1}, #2}
\def \JSQRT #1 #2 {{\em J. Quant. Spectrosc. Radiat. Transfer\/} {\bf #1}, #2}
\def \MN #1 #2 {{\em Mon. Not. R. Astr. Soc.\/} {\bf #1}, #2}
\def \MEM #1 #2 {{\em Mem. R. Astr. Soc.\/} {\bf #1}, #2}
\def \PLR #1 #2 {{\em Phys. Rev. Lett.
\/} {\bf #1}, #2}
\def \PASJ #1 #2 {{\em Publ. Astron. Soc. Japan\/} {\bf #1}, #2}
\def \PASP #1 #2 {{\em Publ. Astr. Soc. Pacific\/} {\bf #1}, #2}
\def \NAT #1 #2 {{\em Nature\/} {\bf #1}, #2}
\def \SAIT #1 #2 {{\em Mem.\ Soc.\ Astron.\ It.\/} {\bf #1}, #2}
\def \MESS #1 #2 {{\em The Messenger\/} {\bf #1}, #2}
\def \ASTRNACH #1 #2 {{\em Astron. Nach.\/} {\bf #1}, #2}

\documentstyle{BSAXPwork}
\input epsf.sty
\begin{opening}
\title{Magnetars in the Afterglow Era }
\author{D. Eichler$^1$, Y. Lyubarsky$^1$, C. Thompson$^2$, P. Woods$^3$}
\institute{$^1$Ben-Gurion University, Beer-Sheva, Israel\\
 $^2$ C.I.T.A., Toronto, Canada \\
$^3$Universities Space Research Association, Huntsville, Alabama }
\date{} 
\end{opening}

\begin{document}

\oddpagefooter{}{}{} 
\evenpagefooter{}{}{} 
\medskip  

\begin{abstract}
The X-ray afterglow that is observed following  large flares on magnetars
can be
accurately fit by
simple and quantitative theoretical models:
The long term afterglow, lasting of order weeks, can be understood  as
thermal radiation
of a heated neutron star crust that is re-scattered in
the magnetosphere.  Short term afterglow is well fit by the cooling
of a non-degenerate, pair-rich layer, which gradually shrinks
and releases heat to a pair-free zone above it.
Measurements of persistent optical and infrared emission directly
probe
long lived currents in the magnetosphere which are a likely
 source of collective plasma emission.  The superstrong
magnetic field plays an important role in generating these various
emissions, and  previous inference of its strength in magnetars is
supported by the good fits with observation.
\end{abstract}

\medskip

\section{Introduction}

  Magnetars exhibit a variety of revealing properties: They generate
soft gamma ray
bursts (SGR events) which typically last $\sim 80 - 200$ ms, and
in extreme cases have super-Eddington luminosities, up to $10^7
L_{\rm Edd}$.
The brightest of these events reach peak luminosity within several milliseconds
to a few seconds, and
have observable tails that taper off within a few seconds to
hundreds of seconds. This prompt burst emission is followed by
short term X-ray afterglows lasting thousands of seconds, and, in
some cases, longer term X-ray afterglows which persist for weeks
or longer. Three AXP's have been discovered to emit optical and
infra-red emission (Hulleman et al. 2000, 2001; Wang \&
Chakrabarty 2002). In one case, the optical emission is pulsed at
the frequency observed in the X-ray bandpass (Kern \& Martin 2002)
and two other AXPs have shown longer timescale variability (of
order weeks) in their IR flux coincident with episodes of burst
activity (Kaspi et al.\ 2002a; Israel et al.\ 2002).

The light curve of some of the largest SGR bursts  has been
explained as a magnetically trapped pair fireball that shrinks to
vanishing size (Thompson \& Duncan 1995; Feroci et al. 2001). Usov
(2001) has proposed that it is due to the cooling of a
 strange quark star surface just after heating by a GRB.
 Long term, transient X-ray afterglow has now been seen on two
occasions from  1900+14, following large flares on Aug. 27, 1998
and April 18, 2001. The longest decay so far has been observed
from 1627-41, lasting about 3 years (Kouveliotou et al., in
preparation). The recent bursts from AXP 2259+586 also show
prolonged X-ray afterglow (Kaspi, Gavriil \& Woods 2002a,b).

Transient afterglow  from neutron star surfaces  following
episodic energy releases  was suggested by Eichler and Cheng
(1989). Such emission could be useful for probing the crust of the
neutron star as well as the depth of the energy release.  There
can be uplifting of surface material following a sufficiently
powerful release of
energy just below the surface, which is then radiated outward
as short term afterglow.
If the total energy release is sufficiently powerful and deep,
then, although  most of the heat is sucked into the star,
some transient afterglow may be observed for weeks or months.
The heat that is absorbed by 
the star reemerges as steady emission over human timescales.

Afterglow radiation also
contains clues about the mechanism of SGR bursts: in particular,
it is sensitive to the temperature to which the magnetosphere is
heated during a burst (Thompson and Duncan 1995) and the manner in
which the rigid crust of the star yields to magnetic stresses
(Lyubarsky, Eichler, and Thompson 2002).
The short term afterglow, with an observed $t^{-0.6}$ power law,
can be explained as the cooling of a pair-supported surface layer
which is heated by exposure to an external fireball, and uplifted
immediately thereafter.
 (Thompson, Woods, Eichler \& Lyubarsky, in preparation).
Heat deposited at depth (or conducted inward) may also contribute
significantly to the persistent X-ray emission of magnetars:
for example, the energy of a giant flare is roughly comparable to the
persistent emission of the SGR integrated over its observed history.

 The nonthermal infra-red and optical emission
may be an important clue about relaxation of the magnetosphere.
Eichler, Gedalin \& Lyubarsky (2002) suggested that the infra-red
and optical emission may be generated by coherent plasma processes
in the magnetospheres of magnetars much as coherent radio emission
is generated in pulsar magnetospheres.

\section{Thermal relaxation of the crust and afterglows.}
A soft gamma ray burst may involve not only a rearrangement of the
magnetic field outside the neutron star, but also  motion,
deformation, and  attendant heating of the crust itself.
The timescale of any afterglow so induced depends on the depth to
which the crust's temperature is significantly raised by the SGR
event.  In order to study cooling of the magnetar crust, let us
assume a deposition of thermal energy density of $\sim 1\times
10^{25}$ erg cm$^{-3}$. This is near the maximum for which
neutrino losses can be neglected, and it is comparable to the
ratio of the flare energy ($\ge 1\times 10^{44}$ ergs) to the
volume of the neutron star. Within the crust,  this energy density
is less than a percent of $B^2/8\pi$, but  {\it greater} than the
pre-existing thermal energy density at depths less than $z_{heat}
\sim 300$ m (for a likely internal temperature of $\sim 5-7\times
10^8 K$; Thompson and Duncan 1996). If deposited over the entire
surface and to a depth of  $\sim 500$ m, this energy density
implies a total energy of a few times the measured Aug. 27
afterglow energy.
A characteristic feature of this heating mechanism is that the
post-burst temperature increases outward in the heated layer, due
to the strong crustal density stratification
and inward heat conduction.

 In our model, which attributes the fading
of the afterglow to the cooling of the magnetar surface, the key
issue is the heat transfer below the surface. The super-strong
magnetic field significantly affects the structure of the upper
crust. The Landau energy is relativistic in a $\sim 10^{15}$ G
magnetic field, $E_L\approx 3B^{1/2}_{15}$ MeV when $B\gg
B_{QED}=4.4\times 10^{13}$ G, and the electron Fermi energy,
$E_F$, becomes comparable with the Landau energy only at a depth
of $\sim 100$ m. At lesser depths electrons are one-dimensional.

 Below a depth of a few meters, the heat is transferred by
degenerate electrons. We calculated the electron thermal
conductivity making use of the code developed by Potekhin (1999).
The electron thermal conductivity, $\kappa$, has a prominent peak
when $E_F$ is about the Landau energy. At larger density, $\kappa$
decreases, reaches a minimum when electrons become effectively
3-dimensional (at $z\sim 2z_1$) and then grows slowly, as in the
nonmagnetized case. At small densities (at $z<z_1$), $\kappa$
rapidly decreases so that close to the surface the heat transfer
is dominated by radiation. Close to the surface, $\kappa$ is so
small that the heat resistance of the crust is dominated by the
upper few meters. The outgoing thermal flux is formed within a
``sensitivity strip'' where the radiation thermal conductivities
become comparable with the electron ones (Gudmundsson et al. 1983;
Ventura \& Potekhin 2001).

 We have developed a
code for simulations of time-dependent, one-dimensional  heat
transfer within the crust of the magnetar.
The calculated outgoing flux is plotted, as a function of time, in
Fig.\ 1 together with the data points obtained by Woods et al.
(2001). The initial temperature distributions in curves 1 through
4 correspond to uniform heat density, with $T$ decreasing inward
until it matched onto the initial (internal) value $T_{int}$. The
heat density was normalized by the temperature $T_{max}$ at the
bottom boundary of the skin zone. The remaining two curves show
that the results are rather robust to varying the initial
conditions. A slight ``knee" occurs when the temperature maximum
passes the minimum of the electron conductivity (at a few times
$10^4$ s for $B = 10^{15}$ G). Beyond this break, the light curve
has a slope which is independent of $B$, because the thermal
conductivity at greater depths approaches the $B=0$ value. An
``ankle'' can occur beyond $10^6$ s, when the temperature maximum
merges with the interior region of almost constant temperature.

\begin{figure}
\epsfysize=20.5cm 
\hspace{1.0cm} \vspace{0.0cm} \epsfbox{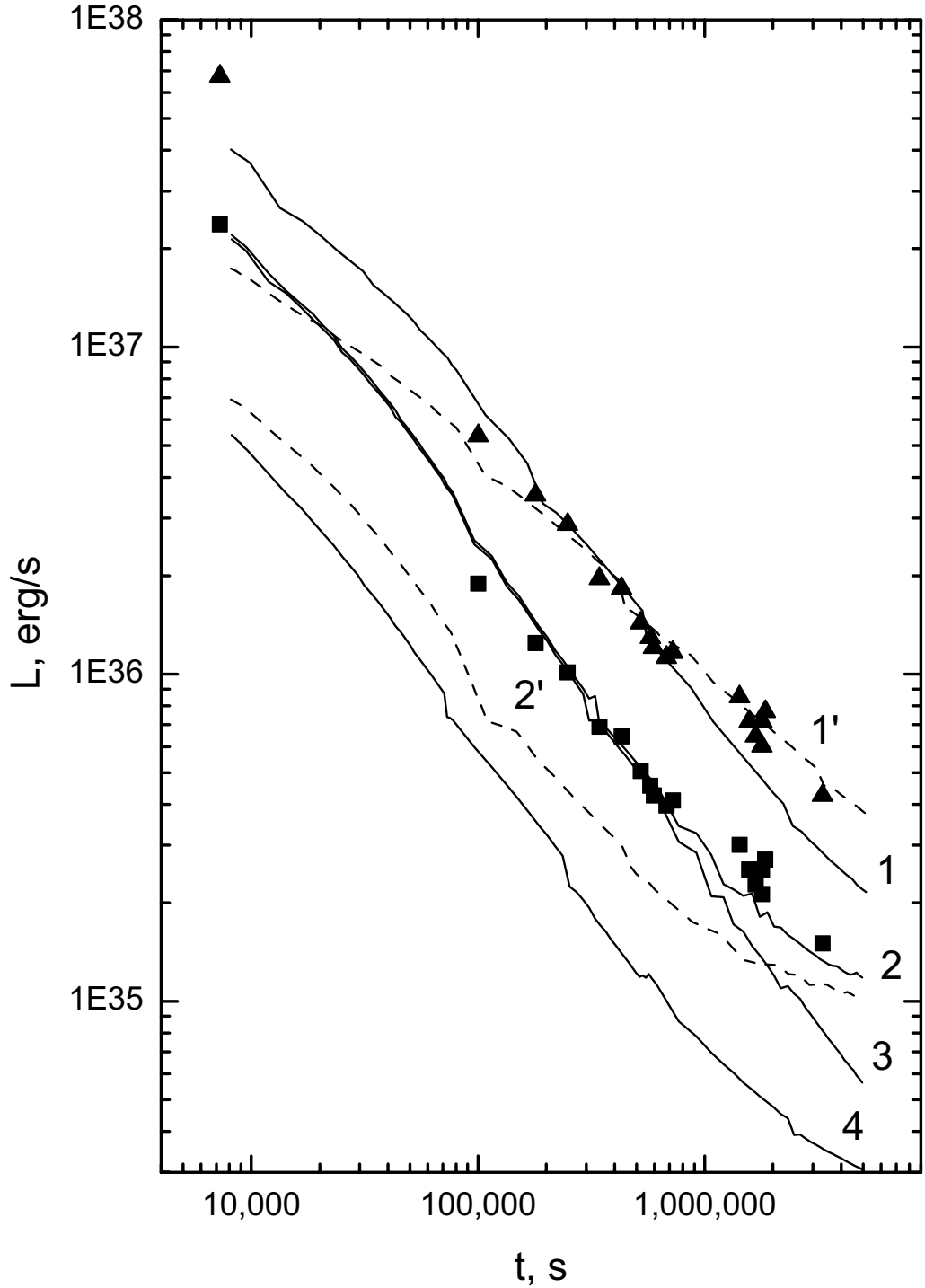}
\caption[h]{Flux times $4\pi\times 10^{12}$ cm$^2$ as a function
of time. Curve 1 corresponds to $T_{max}=5\times 10^9$ K,
$T_{int}=7\times 10^8$ K, $B=10^{15}$ G; curve 2 to
$T_{max}=5\times 10^9$ K, $T_{int}=7\times 10^8$ K, $B=3\times
10^{14}$ G; curve 3 to $T_{max}=5\times 10^9$ K, $T_{int}=4\times
10^8$ K, $B=3\times 10^{14}$ G; and curve 4 to $T_{max}=3\times
10^9$ K, $T_{int}=4\times 10^8$ K, $B=3\times 10^{14}$ G. The
dotted curve 1' is for $B=10^{15}$ G and an initial temperature
distribution $T= 5\times 10^9$ K at $z<30$ m, and $5 \times 10^9K
(z/30{\rm m})^{-0.6}$ at $z>30$ m; curve 2' is for $B= 3\times
10^{14}$G and $T= 5 \times 10^9$ K at $z<100$ m, and proportional
to $z^{-2}$ at greater depths until merging with the internal
temperature of $7 \times 10^8$ K. Data points are from Woods et
al. (2001). Squares are normalized to a distance of 9 kpc for SGR
1900+14 and triangles to 16 kpc.}
\end{figure}

We find that the transient X-ray light curve of SGR 1900+14 in the
40 days following the Aug. 27 event is consistent with the
hypothesis that the SGR is a magnetar made of otherwise normal
material.  While there may be some freedom in choosing the heat
deposition profile, the 40 day timescale is consistent with the
basic physics of an outer crustal layer which is supported by
relativistic degenerate electrons against gravity, and the heat
capacity and conductivity increase considerably with depth. The
power law
 index of the decay, though certainly inconsistent with a constant
 initial temperature,  is found to be weakly sensitive to the
 exact initial temperature profile:  on timescales more than
 a few days, the deeper layers are in any case cooled by inward
 conduction.  Qualitatively, this causes all but
$\sim 20$ percent of the heat to be sucked into the star and
reradiated only over much longer timescales as surface X-ray
emission or neutrinos. The resulting transient afterglow emission
is $\sim 1$ percent of the flare energy, as observed (Woods et al.
2001), if the initial thermal energy density in the crust is
comparable to the ratio of the flare energy to the volume of the
neutron star. This is also consistent with the observation that
the time integrated luminosity of the SGR is dominated by steady
emission rather than by the decaying post-burst flux.

\section{Short Term Afterglow: Cooling of the Pair Supported Atmosphere}
The magnetically trapped fireball heats the surface layers of the
crust and the absorbed heat is reradiated  by surface photon
emission after the the fireball has dissipated. The temperature of
the fireball reaches about 1 MeV for the strong bursts; photons
diffuse into the crust and heat up the surface layer of the depth
about $10^9-10^{10}$ g/cm$^2$ (Thompson \& Duncan 1995). This temperature is
just enough to dissociate nuclei
(including $^4$He).
While the temperature is kept at 1 MeV by heating from above,
the pressure is increased significantly over the initial hydrostatic pressure.
This indicates that, in this conductive layer, the enthalpy per
nucleon decreases with depth. After the hot magnetospheric
plasma dissipates, the heated layer immediately expands, and the
atmosphere is supported by the thermal pressure of the
electron-positron pairs.
The temperature is now low enough ($kT \sim
0.1$-$0.5\,m_ec^2$) that helium quickly recombines, releasing
up to $\sim 7$ MeV per nucleon.  We assume here that
most of the nuclear dissociation energy is restored before radiative
cooling is complete.  (See Thompson et al. 2003 for a more detailed dicussion.)
In this situation, the total
extractable energy per nucleon  $\varepsilon$ is constrained
to a dynamic range of several: it is unlikely to be
more than tens of MeV, in the absence of fine tuning, since otherwise
the material would be blown off the star.
To summarize the above, $\varepsilon$ is likely to lie in the range
of a few to tens of MeV per nucleon, and to decrease slowly with depth.
We normalize
the column of material to an effective Thomson depth $\tau_T = Y_e\sigma_T
\Sigma/m_p = (Y_e\sigma_T/m_p)\int_0^z\rho(z^\prime)dz^\prime$
and choose $\varepsilon(\tau_T) = \varepsilon_0 (\tau_T/\tau_{T,0})^{-\delta}$.

The atmosphere cools by photon diffusion.  This process
may be envisioned as a cooling wave, i.e. an entropy discontinuity at
depth $z^\star(t)$, which propagates inwards. We assume the strong
magnetization provides stability against convection.  A detailed
analysis shows that the pair-rich zone is unstable to the formation
of an annihilation front, above which (at $z \le z^\star$) the medium
is essentially pair free (Thompson et al. 2003). Its specific
enthalpy is well below that of the pair supported
region, so that the flux which enters it from below mostly leaves the
surface, and can therefore be taken to be constant at $z\le z^\star$.
At the front of the wave, the annihilating pairs release the
energy $\varepsilon$ per nucleon, so that this constant flux is just
\begin{equation}\label{flux}
F = \frac{\varepsilon}{m_p} \frac{d\Sigma(z^\star)}{dt}.
\end{equation}
A detailed calculation shows that the temperature above the annihilation
front varies slowly with depth, $T\simeq T_0\,(\tau_T/\tau_{T,0})^\gamma$,
with index $\gamma \simeq {1\over 5}$
and normalization $kT_0 \simeq {1\over 7}m_ec^2$ at $\tau_{T,0} = 10^8$.
Therefore the outgoing flux
varies with time mainly because the column of matter above the cooling
wave front $\Sigma(z^\star)$ increases with time.

In the super strong magnetic field, the radiation energy transfer
is dominated by the extraordinary photons; their free path is very
high and the Rosseland mean scattering cross-section is written as
$\sigma(T,B)=4\pi^2\sigma_T\,(kT/m_ec^2)^2(B_{QED}/B)^2$
(Silant'ev \& Yakovlev 1980). The outgoing flux is
\begin{equation}
F=\frac{m_pc }{3Y_e\sigma(B,T)}\frac{\partial U}{\partial \Sigma}
= {F_0\over \tau_T}\,\left(\frac{B}{B_{QED}}\,\frac{T}{T_0}\right)^2,
\end{equation}
where $U(T)=(1/2)aT^4$ is the energy density of the extraordinary photons,
and $F_0 = acT_0^2(m_ec^2/k_B)^2/12\pi^2$.
 The energy balance equation (1)
is now a differential equation for $\tau_T[z^\star(t)]$.
With the above scaling between $\varepsilon$ and $\tau_T$, one finds
$\tau_T[z^\star(t)]/\tau_{T,0} = [\left({8\over 5}-\delta\right)\,
(Y_e/\tau_{T,0}^2)\,(B/B_{QED})^2\,(F_0\sigma_T/\varepsilon_0)\, t
]^{1/(8/5-\delta)}$, and
\begin{equation}
{F\over F_0} = \left({B\over
B_{QED}}\right)^{10(1-\delta)/(8-5\delta)}\,\tau_{T,0}^{(5\delta-2)/(8-5\delta)}\,
\left[\left({8\over 5}-\delta\right)\,{Y_eF_0\sigma_T\over
\varepsilon_0}\,t\right]^{-3/(8-5\delta)}.
\end{equation}

For $ 0.4\le \delta \le 0.75$, the obtained dependence of the
outgoing flux on time is close to the  overall
 $t^{-0.6}$  dependence  (with slight fluctuations in the spectral index of order $0.1$)
observed from 1900+14 over $10^3$ s following  the Aug 29, 1998
burst (Ibrahim et al. 2001, Lenters et al. 2003). The large
temperature ($kT_{bb} \simeq 4$ keV) of that afterglow at $\sim
10$ s following the burst points to a small radiative area (about
1 percent of the surface area of a neutron star) and a radiative
flux $2 \times 10^{26}(kT_{bb}/4~{\rm keV})^4$ erg cm$^{-2}$
s$^{-1}$.  This implies
\begin{equation}
{B\over B_{QED}} \simeq 10\, \left({\varepsilon_0\over{\rm 10
MeV}}\right)^{-3/4}\, \left({kT_{bb}\over 4~{\rm keV}}\right)^5,
\end{equation}
assuming $\delta =0.6$.

\section{Coherent Emission from Magnetars}
 Magnetars, it has been
proposed (Thompson \& Duncan 1996; Thompson, Lyutikov \& Kulkarni,
2002), have twisted magnetic loops in their magnetospheres.  Most
of the time, the thermal scale height of their atmospheres is too
low to populate the magnetosphere with thermal plasma. On the
other hand, magnetospheric currents can easily be drawn out of the
surface of the star from at least one of the footpoints. A modest
rate of
magnetic field dissipation ($d\ln B/dt\sim 1/10^2$yr) yields a
sufficient potential drop across the length of the loop to create
enough plasma to short out any larger potential drop. The density
of plasma so estimated is many orders of magnitude larger than
that in pulsar magnetospheres. If pulsars can radiate coherently
in the radio, this frequency being ultimately determined by the
plasma frequency in the pulsar magnetosphere, then similar
processes could occur in magnetar magnetospheres with the  plasma
frequency scaled up appropriately.

The total density must clearly be at least as high as the minimum
to deliver the required current,
\begin{equation}
j = \frac{c}{4\pi}|\nabla \times {\bf B}| \simeq ec \times
[2 \times 10^{17} \frac{B_{15}}{R_6} {\rm cm}^{-3}]\sin^2\theta
\Delta\phi_{N-S}.
\end{equation}
Here $\theta$ is the magnetic polar angle and $\Delta\phi_{N-S}$
is the relative twist (in radians) between the north and south
magnetic poles.
Because the above density greatly exceeds the co-rotation charge
density, nearly equal numbers of positive and negative charges are
required to avoid absurdly high electric fields.  Positive charges
may be supplied either by pulling ions off the surface of the
star, or in situ through pair creation.  In the second case, the
particle density can exceed the above by a multiplicity factor
$\eta$ (which may be quite large; e.g. Hibschmann \& Arons 2001).

That a two-species plasma is needed suggests that there is
counterstreaming between the positive and negative charges. This
gives rise to a broad band two stream instability. The excited
plasma waves may be converted into outgoing electromagnetic waves
(see, e.g., Gedalin, Gruman \& Melrose 2002; Lyubarsky 2002).
Regardless of the details of any particular counterstreaming model
for the coherent emission, escaping coherent radiation probably
has a frequency of the order of the plasma frequency in the frame
of the outflowing plasma, which gives a frequency in the observer
frame of $2\omega_p \gamma^{1/2}$, where $\gamma$ is the Lorentz
factor, $\omega_p\equiv (4\pi e^2 n/m_e)^{1/2}$, $n$ the plasma
density in the laboratory frame, and $m_e$ the rest mass of the
electron.

The characteristic Lorentz factor of the plasma may be estimated
as that which will give rise to charges of both signs, which is a
necessary condition for shorting out the strong electric fields
that would otherwise obtain. To create a pair plasma in this
situation, one probably requires
resonant scattering of thermal X-ray photons that emerge from the star's
surface.\footnote{unless the magnetic field is extremely strong,
$B
> 10^{16}$ G, and strongly sheared, in which case a pair corona
can be maintained through multiple non-resonant Compton scattering
(Thompson et al. 2002).} In order to be resonantly scattered by a
relativistic
electron moving in its lowest Landau level, thermal photons of energy
$\epsilon_{\gamma}$ must have frequency of $eB/m_ec$ in the
electron rest frame
-- just as in sub-QED magnetic fields.
Therefore a Lorentz factor of $\gamma \sim
(B/B_{QED})( {m_ec^2}/\epsilon) \sim 10^{3}(10 {\rm keV}/\epsilon)
B_{15}$ is needed.

Making use of the above estimates for the plasma density and Lorentz factor,
one finds the frequency for coherent emission,
\begin{equation}
\nu \sim  \frac 1{\pi}\gamma^{ \frac{1}{2}}\omega_{p}\sim 2 \times
10^{14}\left(\frac{\eta B_{15}\gamma_{3}}{R_6}\sin^2\theta
\Delta\phi_{N-S}\right)^{\frac{1}{2}}\rm Hz.
\end{equation}
Near the surface, where $B_{15}\sim 1$ , this suggests emission in
the near IR, optical, or even UV for high enough $\eta$ (Eichler,
Gedalin and Lyubarsky, 2002). A twisted magnetic arch that
protrudes from a magnetar surface could emit over a broad band,
depending on the exact altitude of the emission.  In the
case where the current is supplied by electrons and ions, Lyutikov (2002)
has suggested that coherent radio emission may be generated.
(We note that strong emission at sub-millimeter wavelengths
is plausible in this case.)

The coherent emission of pulsars is only a small fraction of the
spin-down power, but it can be a much higher fraction of the power
in polar currents, as the latter is itself only a small fraction
of the total. By the same token, a considerable fraction of the
long term magnetic energy dissipation  in magnetars could end up
as coherent electromagnetic emission; 1 to 10 percent is not
unreasonable. A magnetar at a distance of up to 10 kpc could be
detectable at 2.2 microns with imminent technology at a luminosity
of  $10^{33}$ erg/s ( $\sim 10^{-2}$ of its persistent, pulsed
X-ray flux). Such emission would almost certainly have the period
of the magnetar, and would probably be polarized. In analogy to
pulsars, where the direction of polarization can swing with pulse
phase, the time-integrated polarization would probably be less
than that at any instant, but it could nevertheless be non-zero.

Is the optical emission (Hulleman, van Kerkwijk \& Kulkarni 2000)
from the anomalous X-ray pulsar 4U0142+61  coherent?  On the one
hand,  this emission has been reported (Kern \& Martin 2002) to
have the periodicity of the AXP, which suggests that it arises
from near the magnetar. With correction  for reddening, the
optical emission is about
$10^{32}(D/3 Kpc)^2$ erg/s,
and, because the emitting surface is so small, it is almost
certainly non-thermal. On the other hand, general energetic and
thermodynamical considerations still allow incoherent optical
emission from neutron star magnetospheres at detectable levels
(Eichler \& Beskin 2000):  A brightness temperature of up to
$10^{12}(B/G)^{-1/7}\gg 10^6 T_{\odot}$K is allowable in the case
of emission by electrons, and this can in principle provide
detectable optical emission from very small emitting areas. (The
optical pulsar in the Crab exemplifies this.) The lower limit on
the size of the emitting region is likely to come, in the case of
electrons, from the constraints  on the field strength set by the
emission frequency and other considerations.
The frequency and luminosity of the AXP optical/IR emission are also
consistent with incoherent cyclotron radiation from a corona of hot ions
beyond the radius of fast cyclotron cooling (about $30 R_{NS}$;
Thompson et al. 2002).

Overall there are strong reasons to expect a substantial
power in coherent plasma emission from the current-carrying
magnetosphere of an AXP or SGR. Future tests that could conceivably establish
coherence  include a) ultra-fast photometry and polarimetry, which
could reveal rapid time variability (micropulsation); b) pulsed
infrared emission, which could set much higher floors for the
brightness temperature; and c) polarization-time profiles, which
could possibly distinguish between different coherent emission mechanisms.
The peak frequency of this emission provides a strong diagnostic
of the energy and composition of the charge carriers.
\bigskip

In conclusion, soft gamma ray repeaters  and AXP's display a rich
variety of transient,  non-$\gamma$-ray emission that may be
caused or otherwise affected by the SGR events. They seem to be
very nicely explained by the the very same hypotheses that
explained the SGR events themselves: that they are neutron stars
with ultra-strong magnetic fields, that the SGR events are powered
by magnetic energy release that can extend well into the neutron
star crust, and that the magnetospheres are ``twisted'' and support
long-lived currents. That the ultra-strong magnetic field $\sim
10^{15}$ G arises from quantitative fits to the data lends further
support to the claim that the field is indeed so strong.

This research was supported by an Adler Fellowship from the Israel
Science Foundation, the Arnow Chair of Theoretical Astrophysics,
the Israel Ministry of Absorption, and the NSERC of Canada. We
acknowledge helpful conversations with C. Kouveliotou.

\end{document}